\documentstyle[prl,aps,epsfig]{revtex}

\font\mb=msbm10

\begin{document}
\draft
\twocolumn[\hsize\textwidth\columnwidth\hsize\csname@twocolumnfalse\endcsname

\title{\bf The Fractality of the Hydrodynamic Modes of Diffusion}

\author{P. Gaspard and I. Claus\\
{\em Center for Nonlinear Phenomena and Complex Systems}\\
{\em Universit\'e Libre de Bruxelles, Code Postal 231,Campus Plaine, B-1050
Brussels, Belgium}\\
T. Gilbert\\
{\em Department of Chemical Physics,}\\
{\em The Weizmann Institute of Science, Rehovot 76100, Israel}\\
J. R. Dorfman\\
{\em Department of Physics and Institute for Physical Science and
Technology,} \\ {\em University of Maryland, College Park MD, 20742, USA}}
\address{}

\date{\today}

\maketitle

\begin{abstract}
Transport by normal diffusion can be decomposed into the so-called
hydrodynamic modes which relax exponentially
toward the equilibrium state.  In chaotic systems with two degrees of
freedom, the fine scale structure of
these hydrodynamic modes is singular and fractal.  We characterize them by
their Hausdorff dimension which is
given in terms of Ruelle's topological pressure.  For long-wavelength
modes, we derive a striking relation between the Hausdorff dimension,
the diffusion coefficient, and the positive Lyapunov exponent of the
system. This relation is tested numerically on two chaotic systems
exhibiting diffusion, both periodic Lorentz gases, one with hard
repulsive forces, the other with attractive, Yukawa
forces. The agreement of the data with the theory is excellent.
\end{abstract}
]

Recent progress has revealed very interesting relationships between
non-equilibrium statistical mechanics and
and fractal structures that appear in the phase space of
non-equilibrium, chaotic systems.  Theoretical and numerical work has shown
that systems of
interacting particles have a high sensitivity to initial conditions as
characterized by positive Lyapunov
exponents \cite{krylov79,sinai70,livi86,posch88,vanbeijeren95,chernov97}.  As a
consequence, the motion of the particles composing the system is typically
chaotic.  Under stationary or
time-dependent non-equilibrium conditions, this dynamical chaos generates
fractal structures in phase space
\cite{gaspard90,evans90,evmo,hoov,chernov93,tasaki95,gaspbaras95,dorfman95,gaspard98,dorfman99,dettmann00}.  In
this context, Gilbert {\em et al.}\cite{gilbert2000a} have very recently
shown for a multibaker model of
diffusion that the entropy production expected from non-equilibrium
thermodynamics is a consequence of the
fractal character of the hydrodynamic modes.  Moreover, the same authors
have shown that the diffusion
coefficient is related to the fractal dimension of these modes in a family
of multibaker models
\cite{gilbert2000b}.   The important question remains whether such results
can be extended to more realistic systems.

The purpose of this Letter is to describe the fractality of the
hydrodynamic modes of diffusion for general chaotic systems with two
degrees of freedom, including Hamiltonian ones.
The present approach requires neither an escape of particles out of the
system as in the escape-rate formalism,
nor a non-Hamiltonian thermostat as in the thermostatted-system approach.
Both formalisms lead to connections between transport coefficients and
chaotic properties, but in a restrictive framework,
which is not the case here.
We obtain a formula which gives the
Hausdorff dimension of the hydrodynamic modes in terms of: (a) the
dispersion relation of diffusion, and (b) the Ruelle
topological pressure which characterizes the chaotic properties of the
dynamics.  This new formula allows us to
prove that, for
 general chaotic systems with two degrees of freedom, the diffusion
coefficient is given in terms
of the Hausdorff dimension of the modes and the positive Lyapunov exponent,
as conjectured in Ref.
\cite{gilbert2000b}.  The results are numerically tested for two examples
of Hamiltonian Lorentz gases.

The relaxation rate of a hydrodynamic mode of wavenumber ${\bf k}$ is given
by the decay rate of the Van Hove intermediate incoherent scattering function \cite{vanhove} as
\begin{equation}
s_{\bf k} = \lim_{t\to\infty} \frac{1}{t} \ \ln\; \langle \exp\left[ i{\bf
k}\cdot({\bf r}_t-{\bf
r}_0)\right]\rangle \ ,
\label{rate}
\end{equation}
where $\langle\cdot\rangle$ denotes an average over an ensemble of initial
conditions and ${\bf r}$ denotes the
position of the tracer particle.  We assume that the system is mixing so
that this average weakly
converges toward its equilibrium value.  The diffusion coefficient ${\cal
D}$ can be obtained by
expanding the relaxation rate (\ref{rate}) in powers of the wavenumber as
\begin{equation}
s_{\bf k} =-{\cal D} \; {\bf k}^2 + {\cal O}({\bf k}^4) \ ,
\label{dispersion}
\end{equation}
which is known as the dispersion relation of diffusion
\cite{vanbeijeren82}.  In deterministic chaotic systems,
the hydrodynamic modes of diffusion turn out to be given by a singular
distribution without density function
\cite{gaspard96}.  For singular distributions, a cumulative function can
nevertheless be defined by
\begin{equation}
F_{\bf k}(\theta) \equiv \lim_{t\to\infty} \frac{\int_0^{\theta} d\theta'
\exp\left\{ i{\bf k}\cdot \left[
{\bf r}_t(\theta')-{\bf r}_0(\theta')\right]\right\} }{\int_0^{2\pi}
d\theta' \exp\left\{ i{\bf k}\cdot
\left[ {\bf r}_t(\theta')-{\bf r}_0(\theta')\right]\right\} } \ ,
\label{cumul}
\end{equation}
where the initial conditions form a one-dimensional line parameterized by
the angle $\theta\in [0,2\pi[$.  We
suppose that the system has two degrees of freedom so that each trajectory
has one unstable, one
stable, and two neutral directions in the four-dimensional phase space.
The one-dimensional line of initial
conditions is
assumed to be transverse to the unstable directions.
The denominator in Eq. (\ref{cumul}), allows us to define a normalized
cumulative function such that $F_{\bf
k}(0)=0$ and $F_{\bf k}(2\pi)=1$.  The cumulative function forms a curve
$({\rm Re}\; F_{\bf k}, {\rm Im}\;
F_{\bf k})$ in the complex plane $\hbox{\mb C}$. This curve is invariant
under the dynamics and it gives the fine
structure of the hydrodynamic mode of relaxation along the given
one-dimensional line in phase space.

If we assume that the system is chaotic with the Axiom-A properties, we can use
a result by Sinai, Bowen and Ruelle that averages can be performed in terms
of a subset of unstable
trajectories covering the phase space within a certain resolution
\cite{sinai72,bowen75}.  As the time $t$ and
the resolution increase the number of these trajectories also increases so
as to fill densely the phase space in
the limit $t\to\infty$.  The probability weight given to each of these
trajectories is inversely proportional to
their stretching factor $\Lambda_t^{(j)}$ which characterizes its dynamical
instability.  We notice that the
Lyapunov exponent of the trajectory is given by
$\lambda^{(j)}=\lim_{t\to\infty} (1/t)
\ln\vert\Lambda_t^{(j)}\vert$.  Under these circumstances, the average of a
quantity $A$ can be expressed in the
limit $t\to\infty$ as $\langle A \rangle \sim \sum_j
\vert\Lambda_t^{(j)}\vert^{-1} A^{(j)}$, where $A^{(j)}$
is the quantity evaluated for the $j^{\rm th}$ trajectory.  Accordingly,
Eq. (\ref{rate}) can be transformed
into the condition
\begin{equation}
\sum_j \vert\Lambda_t^{(j)}\vert^{-1} \; \exp(-s_{\bf k}t) \; \exp\left\{
i{\bf k}\cdot\left[{\bf
r}_t^{(j)}-{\bf r}_0^{(j)}\right]\right\} \sim_{t\to\infty} 1 \; .
\label{rateSum}
\end{equation}
Indeed, Eq. (\ref{rate}) is obtained after taking the logarithm of Eq.
(\ref{rateSum}), dividing by $t$, and
taking the limit $t\to\infty$.

Now, let us suppose that the sum in Eq. (\ref{rateSum}) is restricted to
the trajectories issued from initial
conditions in the interval $[0,\theta]$.  In this case, we obtain at time
$t$ a polygonal approximation of
the cumulative function Eq. (\ref{cumul}) because, the integral
$\int_0^{\theta}$ is an
average over trajectories with initial conditions in $[0,\theta]$ and the
denominator is proportional to the
factor
$\exp(s_{\bf k}t)$.  In this respect, we notice that Eq. (\ref{rateSum})
with a complete sum is
equivalent to the condition $F_{\bf k}(2\pi)=1$.  Therefore, we can
conclude that, at time $t$, the
curve $({\rm Re}\; F_{\bf k}, {\rm Im}\; F_{\bf k})\subset\hbox{\mb C}$ is
approximated by a
polygon of sides given by the small complex vectors
\begin{equation}
\Delta F^{(j)} = \vert\Lambda_t^{(j)}\vert^{-1} \; \exp(-s_{\bf k}t) \;
\exp\left\{ i{\bf k}\cdot\left[{\bf
r}_t^{(j)}-{\bf r}_0^{(j)}\right]\right\} \; ,
\label{sides}
\end{equation}
as in the construction of fractal curves of von Koch's type \cite{vonkoch}.
Each side has the length
\begin{equation}
\varepsilon_j = \vert\Delta F^{(j)}\vert = \vert\Lambda_t^{(j)}\vert^{-1}
\; \exp(-{\rm Re}\; s_{\bf k}\; t)
\; ,
\label{sizes}
\end{equation}
so that the polygon can be covered by balls of diameter $\varepsilon_j$.
In the limit $t\to\infty$, this polygon converges to a fractal curve
characterized by a Hausdorff dimension
given by $\sum_j \varepsilon_j^{D_{\rm H}}\sim 1$.  Accordingly, the
Hausdorff dimension of
the hydrodynamical mode should satisfy the condition
\begin{equation}
\sum_j \vert\Lambda_t^{(j)}\vert^{-D_{\rm H}} \; \exp(-D_{\rm H}\; {\rm
Re}\; s_{\bf k}\; t) \sim_{t\to\infty} 1
\; .
\label{dimSum}
\end{equation}

On the other hand, Ruelle's topological pressure is defined in dynamical
systems theory by
\cite{gaspbaras95,gaspard98,ruelle78}
\begin{equation}
P(\beta) \equiv \lim_{t\to\infty} \frac{1}{t} \; \ln \; \langle \vert
\Lambda_t\vert^{1-\beta} \rangle \; .
\label{press}
\end{equation}
Since there is no escape of trajectories in our case, we have that
$P(1)=0$.  The mean positive Lyapunov
exponent of the system is given by
$\lambda=-dP/d\beta\vert_{\beta=1}$. Equation (\ref{press}) can be
transformed in the same way as Eq. (\ref{rate}) is transformed into
Eq. (\ref{rateSum}), so that
\begin{equation}
\sum_j \vert\Lambda_t^{(j)}\vert^{-\beta} \; \exp\left[-P(\beta)\; t\right]
\sim_{t\to\infty} 1 \; .
\label{pressSum}
\end{equation}
Comparing with Eq. (\ref{dimSum}), we infer that $\beta=D_{\rm H}$ whereupon
\begin{equation}
P(D_{\rm H}) = D_{\rm H} \; {\rm Re}\; s_{\bf k} \; ,
\label{formula}
\end{equation}
which is the new and central formula of this Letter.  This formula
generalizes Bowen's formula $P(D_{\rm H})=0$
for the Hausdorff dimension of a fractal invariant set of trajectories
\cite{bowen76}.  In the present case, we
are instead dealing with a complex invariant curve associated with a
relaxation process of rate $s_{\rm k}$,
which explains the presence of a new term in the right-hand side of Eq.
(\ref{formula}).  The root of Eq.
(\ref{formula}) gives the Hausdorff dimension $D_{\rm H}({\bf k})$ of the
hydrodynamic mode of wavenumber ${\bf
k}$.

If the wavenumber ${\bf k}$ vanishes, the relaxation rate (\ref{rate}) also
vanishes, $s_{{\bf k}=0}=0$, and Eq.
(\ref{cumul}) shows that the cumulative function becomes $F_{{\bf
k}=0}(\theta)=\theta/(2\pi)$,
which forms a straight line in the complex plane.  In this equilibrium
limit, Eq. (\ref{formula}) reduces to
$P(D_{\rm H})=0$ so that we get the dimension $D_{\rm H}=1$, as it should
be for a straight line.

For a non-vanishing but small wavenumber, the Hausdorff dimension is
expected to deviate from unity.
Inserting $D_{\rm H}=1+\delta$ and the dispersion relation
(\ref{dispersion}) in Eq. (\ref{formula}), we can
expand both sides
in powers of the wavenumber by using the aforementioned properties of
the topological pressure.  This straightforward calculation shows that the
Hausdorff dimension of the
hydrodynamic mode is given by
\begin{equation}
D_{\rm H}({\bf k}) = 1 + \frac{\cal D}{\lambda} \; {\bf k}^2 + {\cal
O}({\bf k}^4) \; ,
\label{dim}
\end{equation}
as previously obtained for the multibaker models \cite{gilbert2000b}.
Accordingly, we have proved the conjecture of
Ref. \cite{gilbert2000b}, that for
general chaotic systems with two degrees of freedom, the
diffusion coefficient can be expressed in terms of the positive Lyapunov
exponent $\lambda$ and the Hausdorff
dimension of the hydrodynamic modes as
\begin{equation}
{\cal D} = \lambda \; \lim_{{\bf k}\to 0} \frac{D_{\rm H}({\bf k})-1}{{\bf
k}^2} \; .
\label{diffdim}
\end{equation}

\begin{figure}
\centerline{\epsfig{file=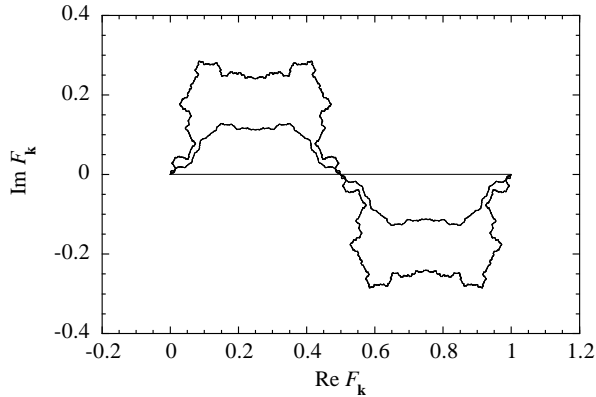,width=8cm}}
\caption{Periodic Lorentz gas where a point particle of unit mass and
velocity undergoes elastic
collisions on hard disks of unit radius forming a triangular lattice with
interdisk distance
$d=2.3$: Curves of the cumulative functions of the hydrodynamic modes of
wavenumber $k_x=0.0$, $0.5$, and
$0.9$ with $k_y=0$.  Note that the fractality increases with the
wavenumber.  The curves are constructed
by averaging over $10^6$ points.}
\label{Fig1}
\end{figure}

\begin{figure}
\centerline{\epsfig{file=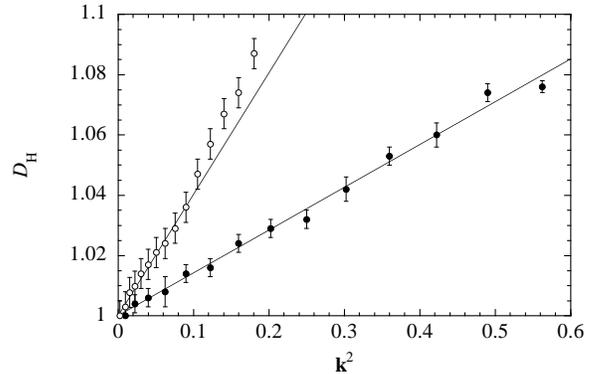,width=8cm}}
\caption{Hausdorff dimension $D_{\rm H}$ of the hydrodynamic modes versus
${\bf k}^2=k_x^2$ ($k_y=0$) for both
periodic Lorentz gases with hard-disk scatterers (filled circles) and
Coulomb scatterers (open circles).
Both solid lines have slopes equal to ${\cal D}/\lambda$ for the respective
diffusion coefficient $\cal D$ and
Lyapunov exponent $\lambda$ of the Lorentz gases.}
\label{Fig2}
\end{figure}

We numerically tested the relation (\ref{diffdim}) for periodic
Lorentz gases, either with repulsive or
with attractive scatterers.
For the case of repulsive scatterers, we considered the periodic Lorentz
gas where a point particle
undergoes elastic collisions on hard disks forming a triangular lattice.
This billiard is known to be fully
chaotic \cite{BS80}.  If the horizon seen by the moving particle is finite
the diffusion coefficient
is positive and finite \cite{BS80,chernov94}, and the higher-order
coefficients such as the super-Burnett
coefficients are finite\cite{chernov2000}.  We take the
radius of the disks to be unity, and the distance between the centers of
the disks
equal to $d=2.3$, so that the finite-horizon condition $2<d<4/\sqrt{3}$ is
satisfied.
Numerical computation shows that the diffusion coefficient is ${\cal
D}=0.25\pm 0.01$ and the positive Lyapunov
exponent $\lambda=1.76\pm 0.01$.  For this system, the curves $({\rm Re}\;
F_{\bf k}, {\rm Im}\; F_{\bf k})$ of
the hydrodynamic modes of wavenumbers $k_x=0.0$, $0.5$, and $0.9$ and
$k_y=0$ are depicted in Fig. \ref{Fig1}.
The initial conditions of the point particle are taken as the
one-dimensional line given in phase space by the
positions ${\bf r}_0(\theta) = (\cos\theta,\sin\theta)$ and the velocities
${\bf v}_0(\theta) =
(\cos\theta,\sin\theta)$.  The Hausdorff dimension of the fractal curves of
Fig. \ref{Fig1} has been computed
by a box-counting algorithm for different values of the wavenumber.  These
values are plotted versus $k_x^2$ in
Fig. \ref{Fig2} (filled circles) together with a solid line of slope ${\cal
D}/\lambda=0.14$.  The very nice
agreement shows that the dimension behaves as expected from Eq.
(\ref{dim}), which provides numerical
evidence of the validity of Eq. (\ref{diffdim}).

\begin{figure}
\centerline{\epsfig{file=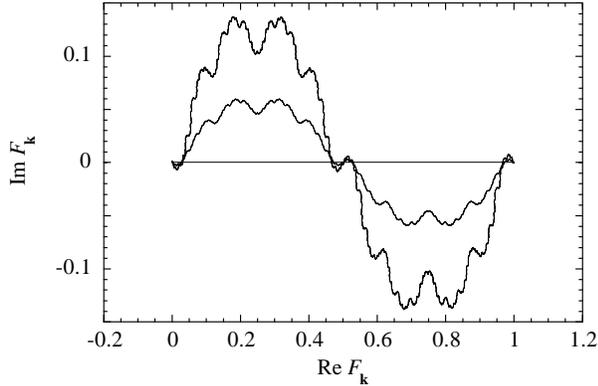,width=8cm}}
\caption{Periodic Lorentz gas where a charged particle of unit mass and
energy $E=3$ moves in a square lattice of
screened Coulomb potentials with $\alpha=2$: Curves of the cumulative
functions of the hydrodynamic modes of
wavenumber $k_x=0.0$, $0.2$, and $0.4$ with $k_y=0$.  Note that here also
the fractality increases with the
wavenumber.  The curves are constructed by averaging over $10^6$ points.}
\label{Fig3}
\end{figure}

For the case of attractive scatterers, we have considered the periodic
Lorentz gas where an electrically
charged particle of mass $m$ moves in a square lattice of screened Coulomb
potentials (also known as Yukawa
potentials).  The Hamiltonian of this system is
\begin{equation}
H = \frac{p_x^2+p_y^2}{2m} - \sum_{\bf l} \frac{\exp(-\alpha \Vert{\bf
r}-{\bf l}\Vert)}{\Vert{\bf r}-{\bf
l}\Vert} \; ,
\label{Yukawa}
\end{equation}
where ${\bf r}=(x,y)$, ${\bf l}=(l_x,l_y)$ with $l_x,l_y=0,\pm 1,\pm 2,\pm
3,...$ and $\Vert\cdot\Vert$ is the
Euclidean distance.  This system was  proved by Knauf to be fully chaotic
and diffusive if the energy of the
moving particle is positive and large enough \cite{knauf}.  We have here
taken the parameter values $m=1$,
$\alpha=2$, and the energy
$E=3$.  We have numerically computed the diffusion coefficient ${\cal
D}=2.5\pm 0.1$ and the Lyapunov exponent
$\lambda=6.2\pm 0.1$.  The curves $({\rm Re}\; F_{\bf k}, {\rm Im}\; F_{\bf
k})$ of the hydrodynamic modes of
wavenumbers $k_x=0.0$, $0.2$, and $0.4$ and $k_y=0$ are depicted in Fig.
\ref{Fig3}.  The initial conditions of
the point particle are taken on a circle centered around the scatterer at
$x=y=0$, with positions ${\bf
r}_0(\theta) = 0.25 (\cos\theta,\sin\theta)$ and velocities ${\bf
v}_0(\theta) = v_0 (\cos\theta,\sin\theta)$
such that $E=3$.  The Hausdorff dimension of these fractal curves are also
plotted in Fig. \ref{Fig2} (open
circles), now together with a solid line of slope ${\cal D}/\lambda=0.40$.
Here again, we observe a good
agreement between both the numerical data and the theoretical prediction of
Eq. (\ref{dim}).  In this case, the
data points deviate from the solid line at large values of $k_x^2$.  Such a
behavior is due to the terms of
order ${\bf k}^4$ and higher in Eq. (\ref{dim}).  These terms turn out to
be more important in the case of the
Coulomb scatterers than in the case of the hard-disk scatterers.
Nevertheless, the data points converge to the
unit value tangentially to the solid line, as predicted by Eq. (\ref{dim}).
Therefore, we have here also a good
confirmation of Eq. (\ref{diffdim}) for the Hamiltonian system (\ref{Yukawa}).

In conclusion, we have derived the formula (\ref{formula}) for the
Hausdorff dimension of
the hydrodynamic modes of diffusion in chaotic systems with two degrees of
freedom.  This formula generalizes
Bowen's formula and it allows us to establish in more generality a relation
between the diffusion
coefficient, the Lyapunov exponent, and the Hausdorff dimension of the
hydrodynamic modes as conjectured in
Ref. \cite{gilbert2000b}.  Our results exhibit new
relations between transport coefficients
and chaotic properties for systems
relaxing to equilibrium, extending those previously obtained in the escape-rate
formalism \cite{gaspard90,gaspbaras95,dorfman95} and in the Gaussian
thermostat method \cite{evans90,evmo,hoov,chernov93,dettmann00}. Therefore,
they illustrate the great importance
of such relationships for both Hamiltonian and non-Hamiltonian systems.
Extensions of these and related results
to fluid systems are now under study by the present authors.

We thank C.~Dettmann, G.~Nicolis, S.~Tasaki, and Howard Weiss for
discussions.  JRD thanks the National Science
Foundation for support under grant PHY 98-20824.  TG is a Feinberg
Postdoctoral Fellow.  IC and PG thank the FNRS
Belgium and the IAP Program of the Belgian Federal OSTC  for financial support.


\begin{thebibliography}{99}

\bibitem{krylov79} N.~N. Krylov, Nature {\bf 153}, 709 (1944); N.~N.
Krylov, {\em
Works on the Foundations of Statistical Mechanics} (Princeton University
Press, 1979);
Ya.~G. Sinai, {\em ibid.} p. 239.

\bibitem{sinai70} Ya. G. Sinai, Russian Math. Surveys {\bf 25}, 137 (1970).

\bibitem{livi86} R. Livi, A. Politi, and S. Ruffo, J. Phys. A: Math. Gen.
{\bf 19},
2033 (1986).

\bibitem{posch88} H. A. Posch and W. G. Hoover, Phys. Rev. A {\bf 38}, 473
(1988);
Ch. Dellago, H. A. Posch, and W. G. Hoover, Phys. Rev. E {\bf 53}, 1485 (1996).

\bibitem{vanbeijeren95}
H. van Beijeren and J.~R. Dorfman, Phys. Rev. Lett. {\bf 74}, 4412 (1995);
{\bf 76}, 3238 (1996); H. van Beijeren, J. R. Dorfman, H. A. Posch, and Ch.
Dellago,
Phys. Rev. E {\bf 56}, 5272 (1997); R. van Zon, H. van Beijeren, and Ch.
Dellago, Phys. Rev. Lett. {\bf
80}, 2035 (1998).

\bibitem{chernov97} N. I. Chernov, J. Stat. Phys. {\bf 88}, 1 (1997).

\bibitem{gaspard90} P. Gaspard and G. Nicolis, Phys. Rev. Lett. {\bf 65},
1693 (1990).

\bibitem{evans90} D. J. Evans, E. G. D. Cohen, and G. Morriss, Phys. Rev. A
{\bf 42},
5990 (1990).

\bibitem{evmo} D. J. Evans and G. Morriss, {\em Statistical Mechanics of
Nonequilibrium Liquids}, (Academic Press, London, 1990).

\bibitem{hoov} W.~Hoover, {\em Computational Statistical Mechanics},
(Elsevier Science Publishers, Amsterdam, 1991).

\bibitem{chernov93} N. I. Chernov, G. L. Eyink, J. L. Lebowitz, and Ya. G.
Sinai, Phys. Rev. Lett. {\bf 70},
2209 (1993); Commun. Math. Phys. {\bf 154}, 569 (1993).

\bibitem{tasaki95} S. Tasaki and P. Gaspard, J. Stat. Phys. {\bf 81}, 935
(1995).

\bibitem{gaspbaras95} P. Gaspard and F. Baras, Phys. Rev. E {\bf 51}, 5332
(1995).

\bibitem{dorfman95} J. R. Dorfman and P. Gaspard, Phys. Rev. E {\bf
51}, 28 (1995); P. Gaspard and J. R. Dorfman, Phys. Rev. E {\bf 52}, 3525
(1995).

\bibitem{gaspard98}
P.~Gaspard, {\em Chaos, Scattering, and Statistical Mechanics}
(Cambridge University Press, Cambridge, 1998).

\bibitem{dorfman99}
J.~R.~Dorfman, {\em An Introduction to Chaos in Nonequilibrium Statistical
Mechanics}
(Cambridge University Press, Cambridge UK, 1999).

\bibitem{dettmann00}
C.~P. Dettmann, {\em The Lorentz gas as a paradigm for nonequilibrium
stationary states}; and T. T\'el and J. Vollmer, {\em Entropy balance,
multibaker maps, and the dynamics of the Lorentz gas},
in:  D. Szasz, Editor, {\em Hard Ball Systems and Lorentz Gas}, Encycl.
Math. Sci.
(Springer, Berlin, 2000).

\bibitem{gilbert2000a} T. Gilbert, J. R. Dorfman, and P. Gaspard, Phys.
Rev. Lett.
{\bf 85}, 1606 (2000).

\bibitem{gilbert2000b} T. Gilbert, J. R. Dorfman, and P. Gaspard, {\em
Fractal dimensions of
the hydrodynamic modes of diffusion}, (preprint {\tt nlin.CD/0007008},
server {\tt
xxx.lanl.gov}).

\bibitem{vanhove} L. Van Hove, Phys. Rev. {\bf 95}, 249 (1954).

(Dover, New York, 1980).

\bibitem{vanbeijeren82} H. van Beijeren, Rev. Mod. Phys. {\bf 54}, 195 (1982).

\bibitem{gaspard96} P.~Gaspard, Phys. Rev. E {\bf 53}, 4379 (1996).

\bibitem{sinai72} Ya. G. Sinai, Russian Math. Surveys {\bf 27}, 21 (1972).

\bibitem{bowen75} R. Bowen and D. Ruelle, Invent. Math. {\bf 29}, 181 (1975).

\bibitem{vonkoch} H. von Koch,
translated in: {\em Classics on Fractals}, G. A. Edgar, Ed.
(Addison-Wesley, Reading MA, 1993); H. von Koch, Acta
Mathematica {\bf 30}, 145 (1906).

\bibitem{ruelle78} D. Ruelle, {\em Thermodynamic Formalism}
(Addison-Wesley, Reading MA, 1978).

\bibitem{bowen76} R. Bowen, Publ. Math. IHES {\bf 50}, 11 (1976).

\bibitem{BS80} L. A. Bunimovich and Ya. G. Sinai, Commun. Math. Phys. {\bf
78}, 247, 479 (1980).

\bibitem{chernov94} N. I. Chernov, J. Stat. Phys. {\bf 74}, 11 (1994).

\bibitem{chernov2000} N. I. Chernov and C. P. Dettmann, Physica A {\bf
279}, 37 (2000);
C. P. Dettmann, {\it The Burnett expansion of the periodic Lorentz gas}
(preprint {\tt nlin.CD/0003038}, server {\tt xxx.lanl.gov}).

\bibitem{knauf} A. Knauf, Commun. Math. Phys. {\bf 110}, 89 (1987); Ann.
Phys. (N. Y.) {\bf 191}, 205 (1989).






\end{thebibliography}
\end{document}